\documentclass[twocolumn,showpacs,preprintnumbers,amsmath,amssymb,superscriptaddress]{revtex4-1}
\usepackage{graphicx}
\usepackage{dcolumn}
\usepackage{bm}
\usepackage{float}

\begin{document}

\selectfont

\title{Inverse targeting - an effective immunization strategy}

\author{Christian M. Schneider}
\email{schnechr@ethz.ch}
\affiliation{Computational Physics, IfB, ETH Zurich, Schafmattstrasse 6, 8093 Zurich, Switzerland}
\affiliation{Department of Civil and Environmental Engineering, MIT, 77 Massachusetts Avenue, Cambridge, MA 02139, USA}
\author{Tamara Mihaljev}
\affiliation{Computational Physics, IfB, ETH Zurich, Schafmattstrasse 6, 8093 Zurich, Switzerland}
\author{Hans J. Herrmann}
\affiliation{Computational Physics, IfB, ETH Zurich, Schafmattstrasse 6, 8093 Zurich, Switzerland}
\affiliation{Departamento de F\'{\i}sica, Universidade Federal do Cear\'a, 60451-970 Fortaleza, Cear\'a, Brazil}

\date{\today}

\begin{abstract}
{
We propose a new method to immunize populations or computer networks against epidemics which is more efficient than any method considered before. The novelty of our method resides in the way of determining the immunization targets. First we identify those individuals or computers that contribute the least to the disease spreading measured through their contribution to the size of the largest connected cluster in the social or a computer network. The immunization process follows the list of identified individuals or computers in inverse order, immunizing first those which are most relevant for the epidemic spreading. We have applied our immunization strategy to several model networks and two real networks, the Internet and the collaboration network of high energy physicists. We find that our new immunization strategy is in the case of model networks up to $14\%$, and for real networks up to $33\%$ more efficient than immunizing dynamically the most connected nodes in a network. Our strategy is also numerically efficient and can therefore be applied to large systems.}
\end{abstract}



\maketitle

\section{Introduction}
The threat of global spreading of epidemics, like the pandemic flu from 2009, or spreading of computer viruses which are endangering the functioning of Internet dependent facilities, are responsible for the enormous increase in public interest on immunization during the last years. Much progress has been achieved in understanding epidemic spreading \cite{socialnetworks,NewmanEpidemicsSpread}, and various models have been developed suggesting possible ways of efficient immunization \cite{holmeAttacks,holmeImmunization,shlomoImmunization,christianArxiv,shlomoEGP,havlin2,callaway,barab}. However, the search for even more effective immunization strategies must be pursued since any improvement of immunization efficiency can save human lives and resources. In this paper we introduce a novel immunization strategy based on inverse targeting, which proves to be effective and numerically more efficient than any proposed previous one.

Epidemics can spread in human population through networks of social contacts, and viruses can propagate on computer networks. We will implement an immunization process on networks by immunizing the nodes which represent either people that are to be vaccinated, or computers that should be equipped with specially developed antivirus programs. 

We suppose that the dynamics of both, the epidemic spreading and the immunization process are much faster than the growth and change of the network itself. The question we want to answer is, given the topology of the network through which the epidemics can spread, what is the best possible way to immunize its nodes. The immunization process should require the smallest possible number of immunization doses. At the same time, the size of the largest connected cluster of non-immunized nodes should stay as small as possible throughout the immunization process.
\begin{figure*}[t]\centering
 \includegraphics[width=2.7cm,angle = 0]{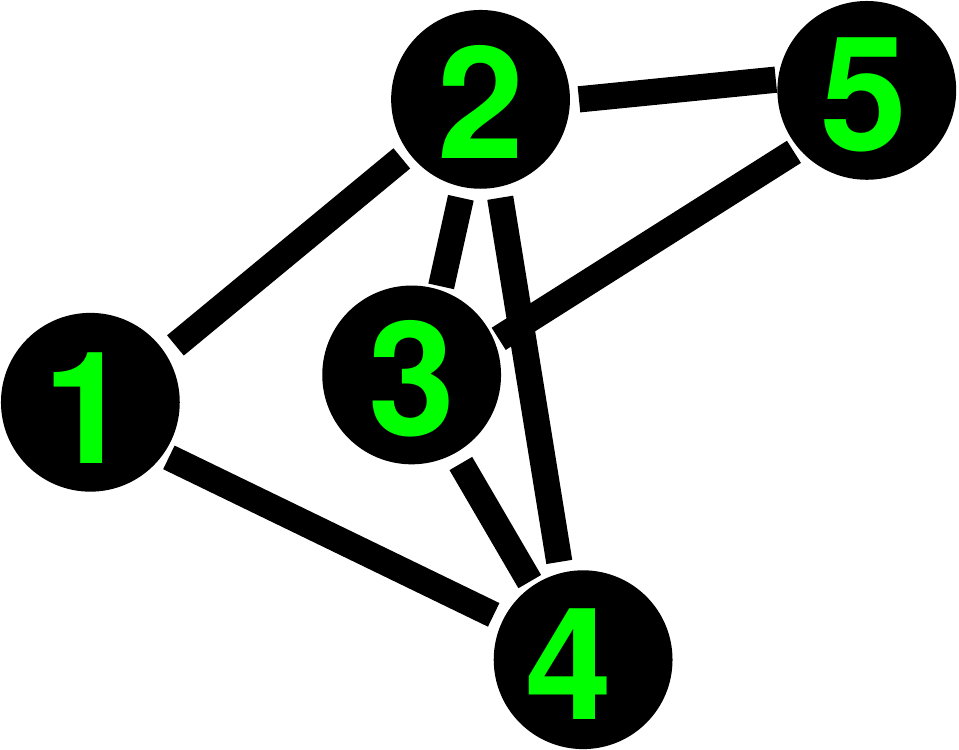}
 \hspace{0.1cm}
 \includegraphics[width=2.7cm,angle = 0]{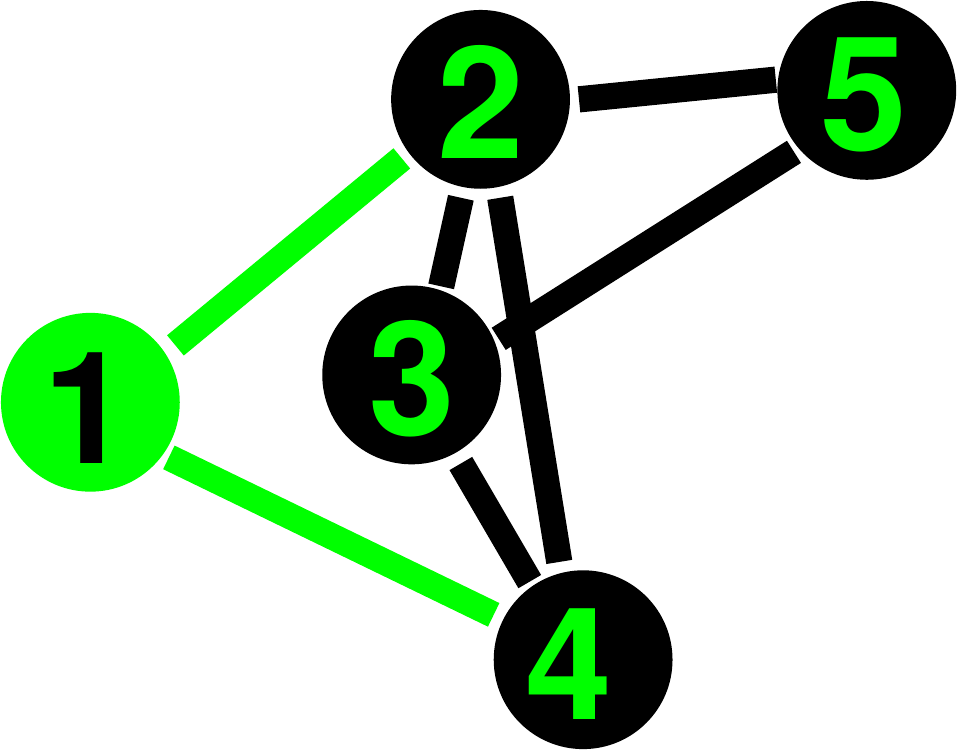}
 \hspace{0.1cm}
 \includegraphics[width=2.7cm,angle = 0]{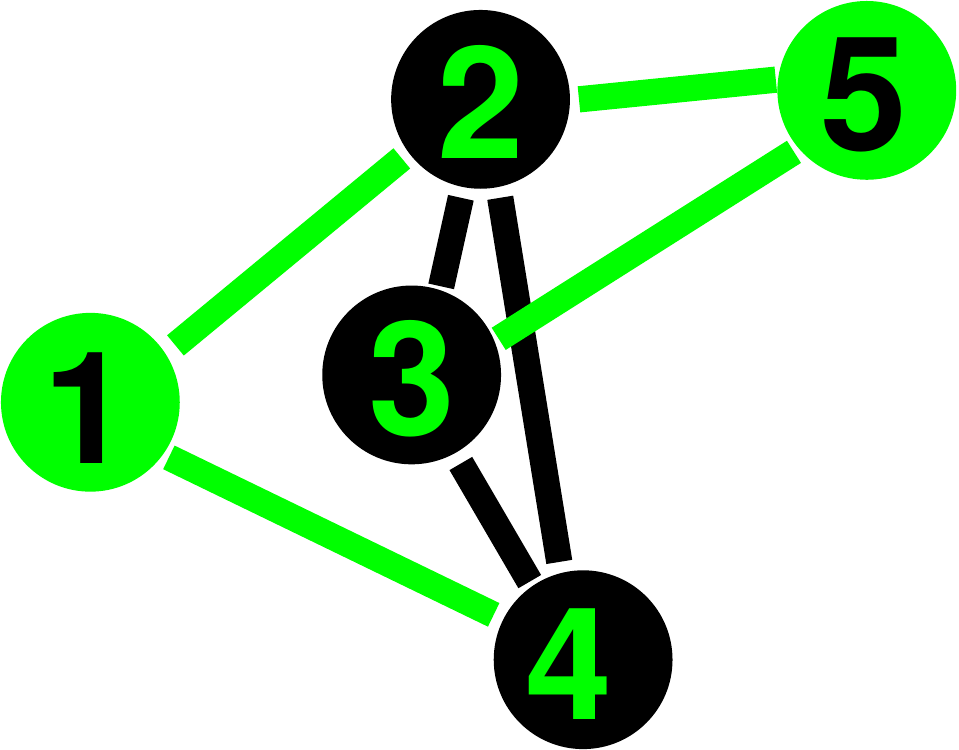}
 \hspace{0.1cm}
 \includegraphics[width=2.7cm,angle = 0]{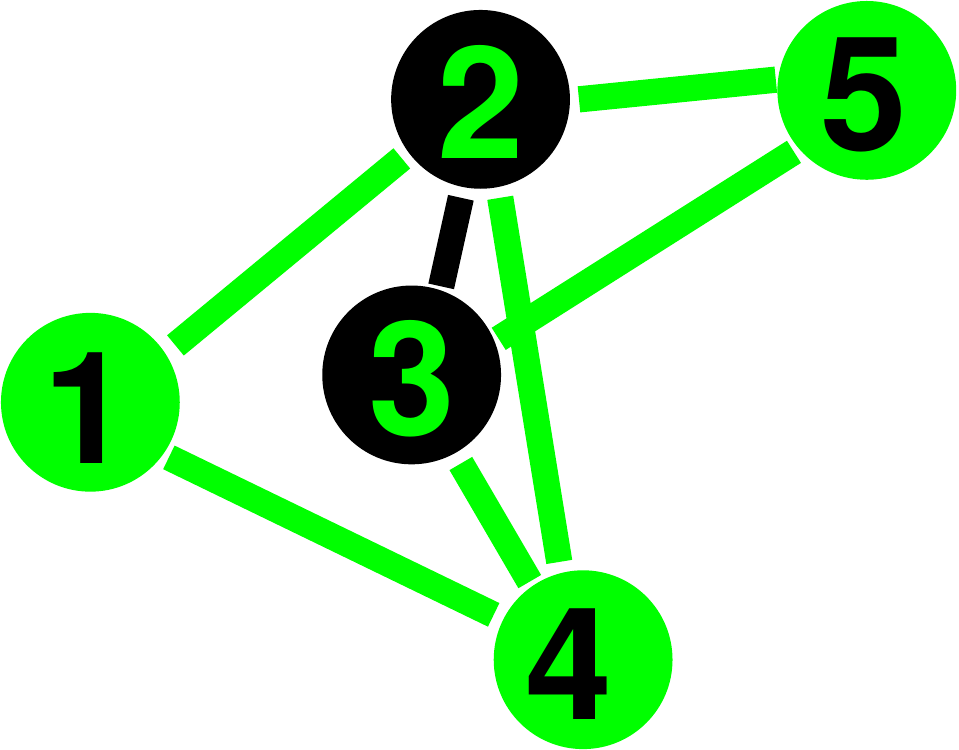}
 \hspace{0.1cm}
 \includegraphics[width=2.7cm,angle = 0]{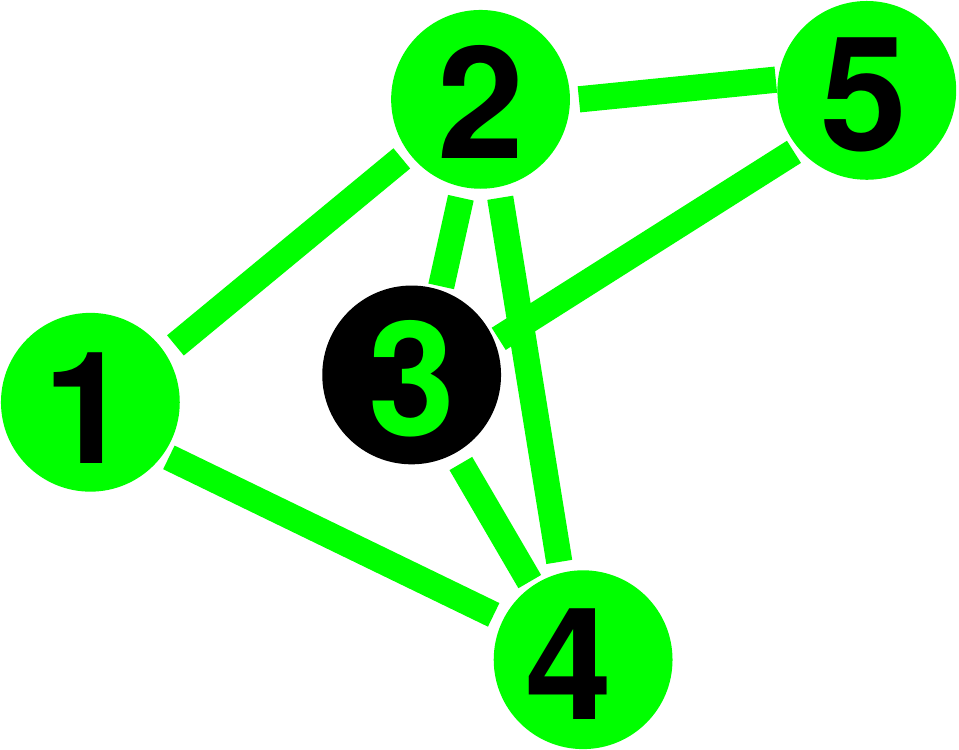}
 \hspace{0.1cm}
 \includegraphics[width=2.7cm,angle = 0]{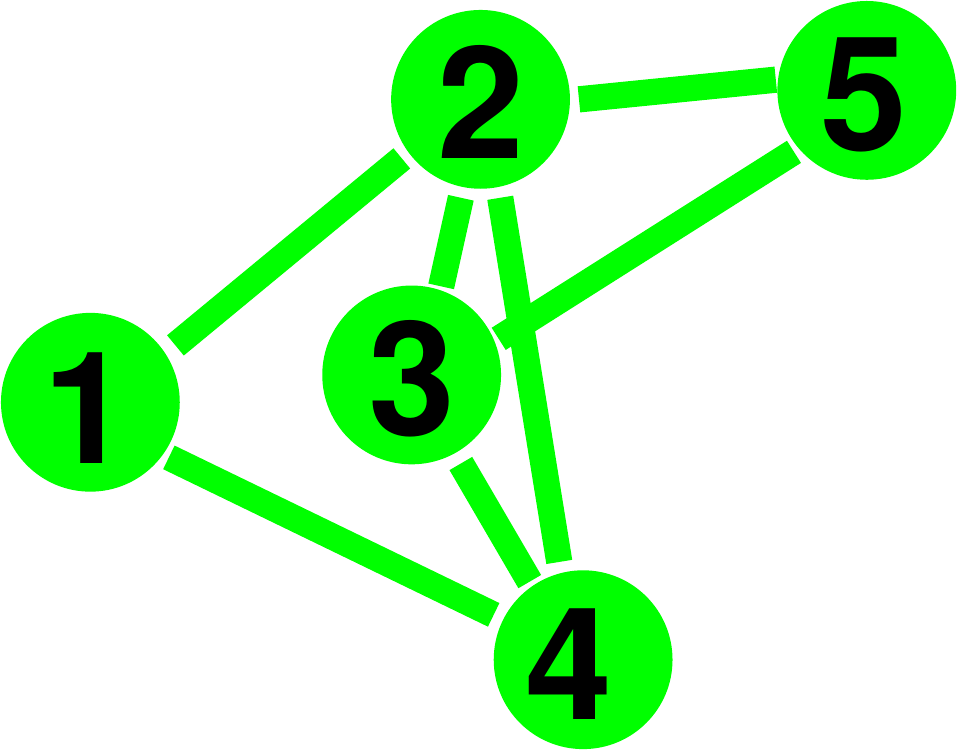}\\~\\
 \includegraphics[width=14cm,angle = 0]{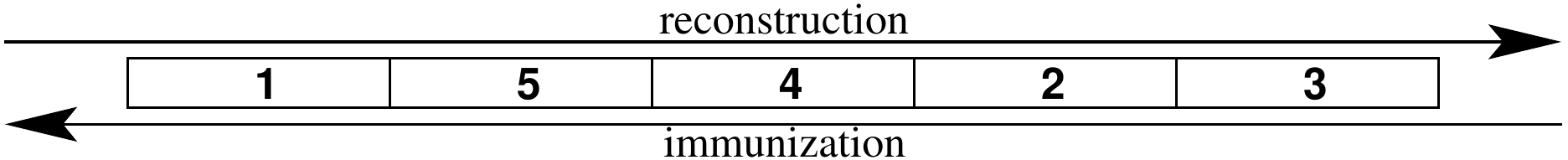}
 \caption{(Colors online) Illustration of the inverse targeting on a very small network. In each step of the reconstruction process (from left to right) a black node is chosen which when colored green together with its links, gives the smallest contribution to the size of the largest green cluster. The inverse of the sequence of nodes obtained from the reconstruction process defines the immunization strategy. In each step of the immunization process (from right to left) a node will be immunized following this inverse sequence.}
\label{fig:topology}
\end{figure*}

An immunization strategy determines the sequence in which nodes of a network should be immunized to prevent most efficiently the spreading of epidemics. The simplest immunization strategy is the one in which the nodes are chosen at random \cite{havlin2}. This strategy has however proven to be inefficient, requiring a large fraction of nodes (in many cases, such as Internet, nearly 100\%) to be immunized in order to stop epidemics spreading \cite{epidemicsSF, InternetBreakdown}. Targeted immunization strategies turn out to be much more efficient \cite{randomVstargeted}. Here one immunizes first those individuals that are most important for the disease spreading. In terms of network properties this importance can be expressed through the number of links of a node, or through its betweenness centrality \cite{betweennessDef,NewmanBetwDef}. Although it has been shown that these two properties are correlated \cite{holmeAttacks}, their role in the immunization process is typically quite different. Most studied targeted strategies are based on these two node traits. 

In the high degree based (HD)\cite{barab,callaway}, and the high betweenness based (HB) \cite{holmeAttacks} targeting the sequence in which nodes will be immunized is based on their degree or betweenness in the initial network. They can be improved by adaptive strategies \cite{holmeAttacks}, high degree adaptive (HDA), and high betweenness adaptive (HBA) strategy. Here the degree or betweenness of nodes is recalculated for the remaining network of non-immunized nodes each time a node has been immunized. The node with highest value of this recalculated degree (for HDA) or betweenness (for HBA) is immunized next. These dynamic strategies have proven to be very effective \cite{holmeAttacks}. 

Other strategies, like immunizing neighbors of a randomly chosen node according to some rule \cite{holmeImmunization,shlomoImmunization} have the advantage of requiring only local information, but are also less effective.
A further approach, recently introduced in Ref. \cite{christianArxiv}, is the improvement of existing strategies through optimization procedures. This approach leads to very effective node and link immunization strategies.
Another strategy based on equal graph partitioning \cite{shlomoEGP} is especially effective when the number of available immunization doses is fixed. For any given number of nodes this strategy finds an effective way to partition the network in clusters such that the size of the largest cluster is the smallest possible. In Ref. \cite{shlomoEGP} the authors show that for a given number of doses this strategy is more effective than targeting HD, HB or HDA strategy. However, if the supply of immunization doses is not known in advance or increases with time, one must take into account that when new doses become available the network of non-immunized nodes can be quite different from the initial network. The nodes which can now be immunized are then different from those that would give an ideal graph partitioning if we had this larger quantity of immunization doses from the beginning of the process.

In such cases dynamic HDA remains the best known immunization strategy because the HBA, which needs more global information, turn out to be numerically extremely demanding. As a consequence, although the HBA may be theoretically the more efficient strategy \cite{holmeAttacks}, the complexity of its algorithm makes it uninteresting for practical purposes and limits its use to very small systems. In this paper we will introduce a strategy which is more efficient than HDA and in many cases even better than HBA, being at the same time numerically much faster than HBA allowing to apply it to large, real systems.
\begin{figure*}\centering
\includegraphics[width=4.2cm, angle= -90]{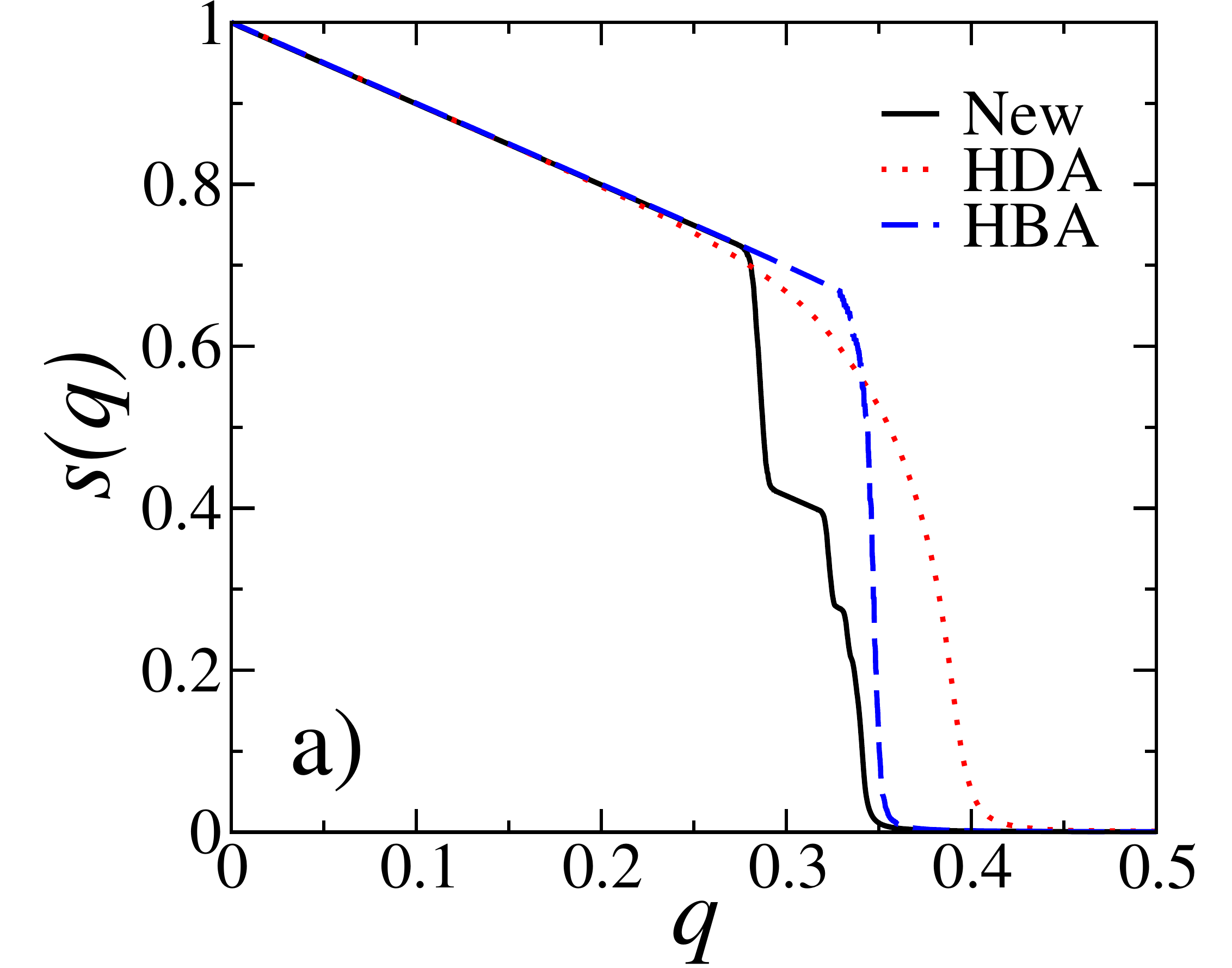}
\includegraphics[width=4.2cm, angle= -90]{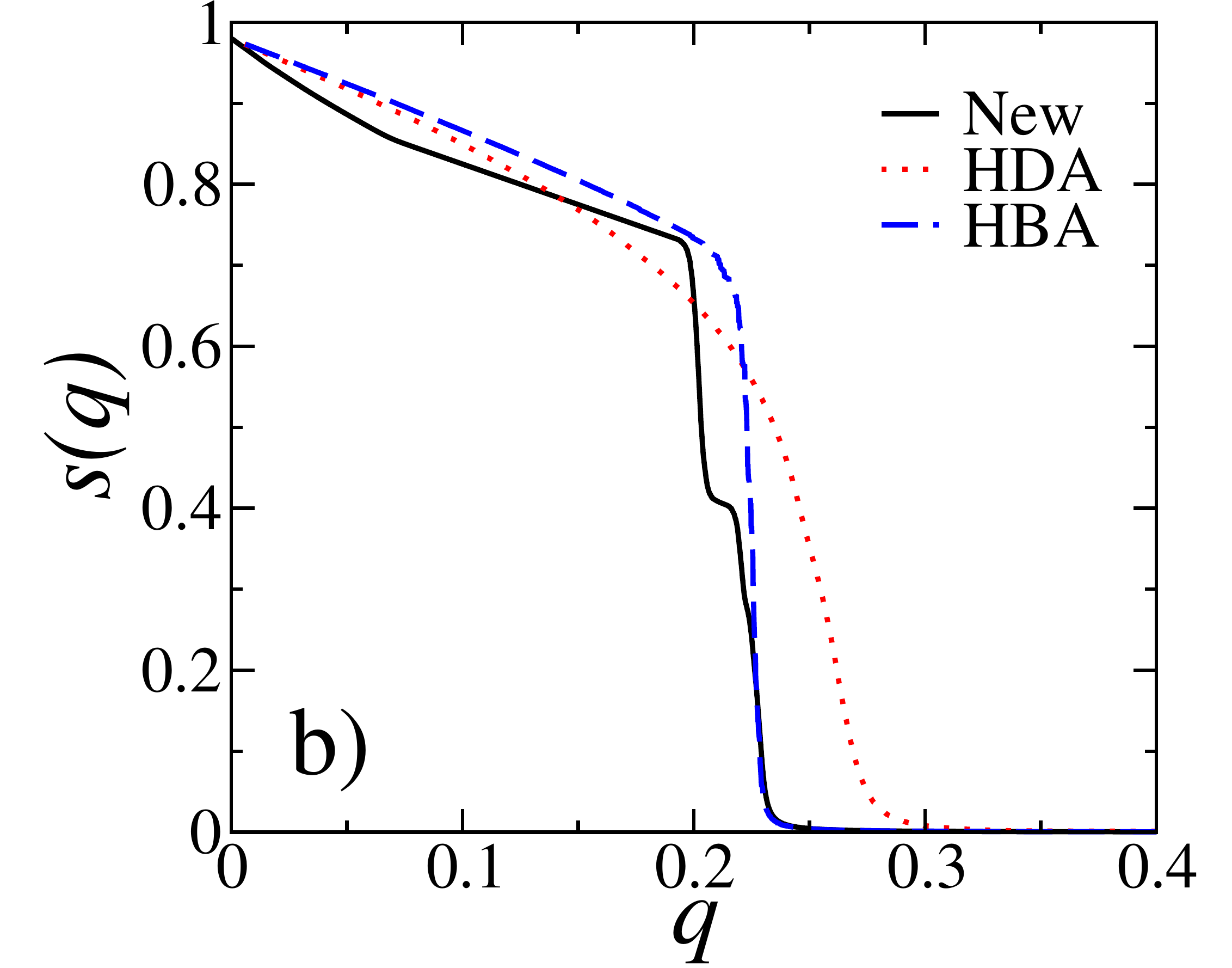}
\includegraphics[width=4.2cm, angle= -90]{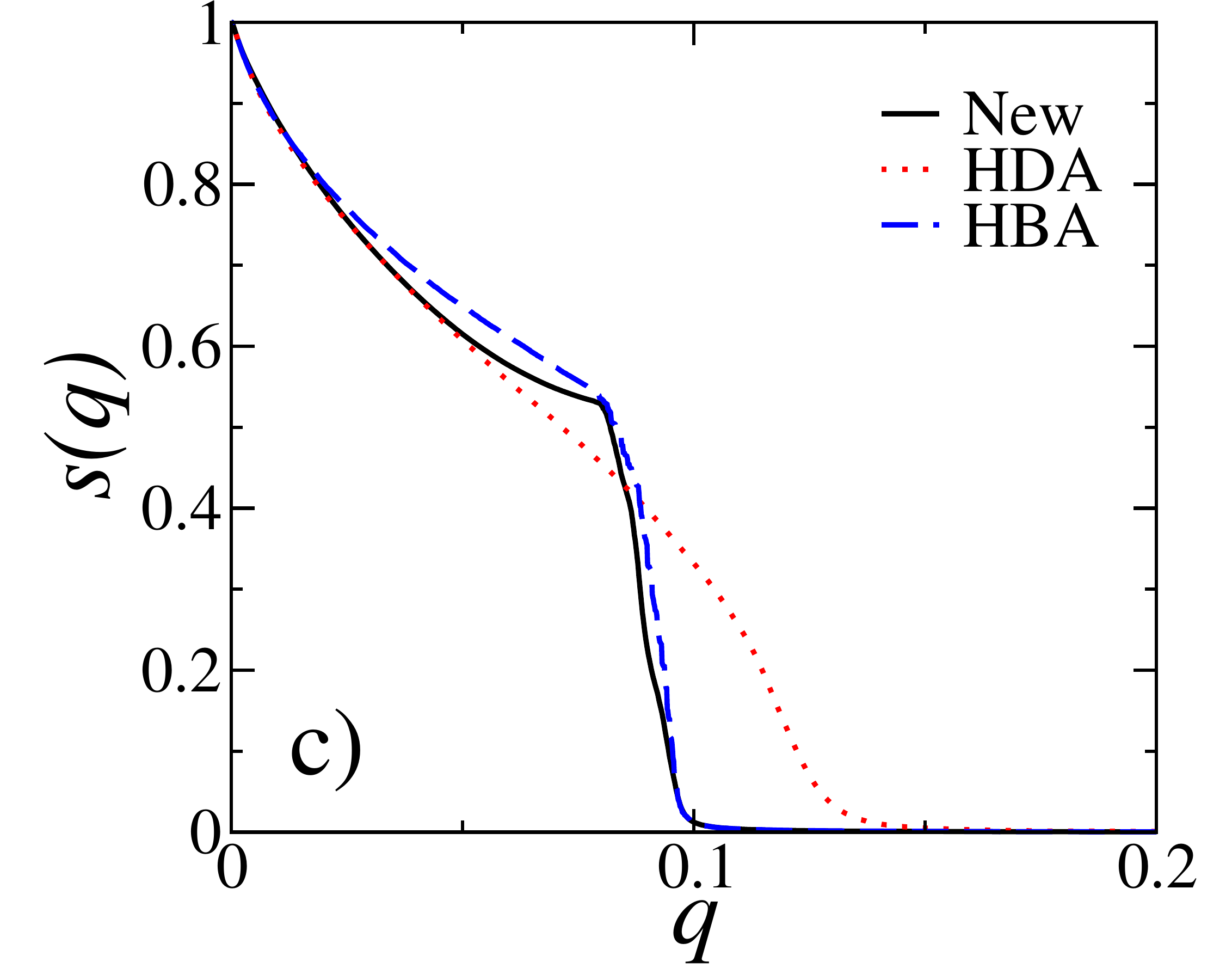}
\caption{The fraction $s(q)$ of sites belonging to the largest connected cluster versus the fraction $q = Q/N$ of immunized nodes using the new immunization strategy (green dashed lines), the HDA attacks (red doted lines) and HDA (black full lines) for (a) random regular networks with $N = 8000, M = 16000$, (b) Erd\H{o}s-R\'enyi networks with $N = 8000, M = 16000$ and (c) scale-free networks with $N = 8000$ and $\gamma = 2.5$~.}
\label{fig:robustness}
\end{figure*}

\section{Inverse targeting}
In any successful immunization strategy, nodes which are not relevant for keeping the largest cluster of non-immunized nodes together, should only be immunized during later stages. In our method we start by recognizing such nodes first. This way we reveal the sequence of nodes that are going to be immunized in reversed order. To illustrate our method, let the nodes and the links of the network be black. Every node has an index (ID) (Fig.\ref{fig:topology}). At each step we check for each black node how would turning it and its links green increase the size of the largest connected green cluster and then choose the one with the smallest contribution to its size. When more nodes give the same contribution, the one with the smallest number of links connecting it to black nodes is chosen. If there are several nodes which fulfill both criteria, one of them is chosen at random. After a node is chosen we put its ID in a list and continue the process until all nodes and links are colored green. The list of nodes' IDs describing the reconstruction process gives us the inverse sequence of nodes that are to be immunized.

The complexity of the inverse targeting algorithm is at most $O(MN)$, or $O(N^2)$ for sparse networks, where $M$ is the number of links and $N$ the number of nodes. This is much faster then the HBA strategy. The best known algorithm for determining betweenness of nodes goes as $O(MN)$ \cite{NewmanBetweenness,BrandesBetweenness} and since it has to be applied at each step of the immunization process, the complexity of the algorithm for the HBA strategy is $O(MN^2)$, or $O(N^3)$ for sparse networks.
\begin{figure}[htb!]\centering
\includegraphics[width=4.2cm, angle= -90]{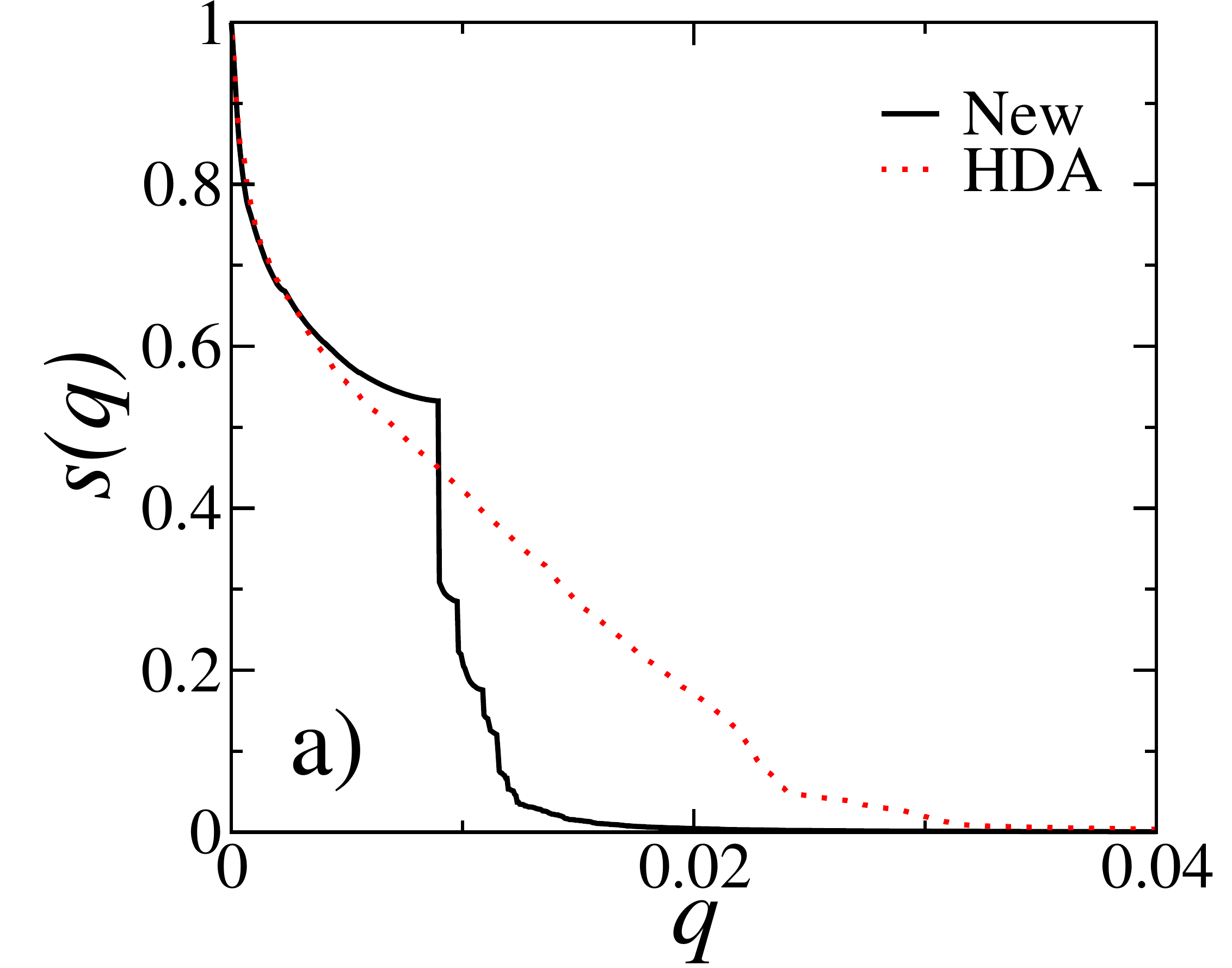}
\hspace{-40px}
\includegraphics[width=4.2cm, angle= -90]{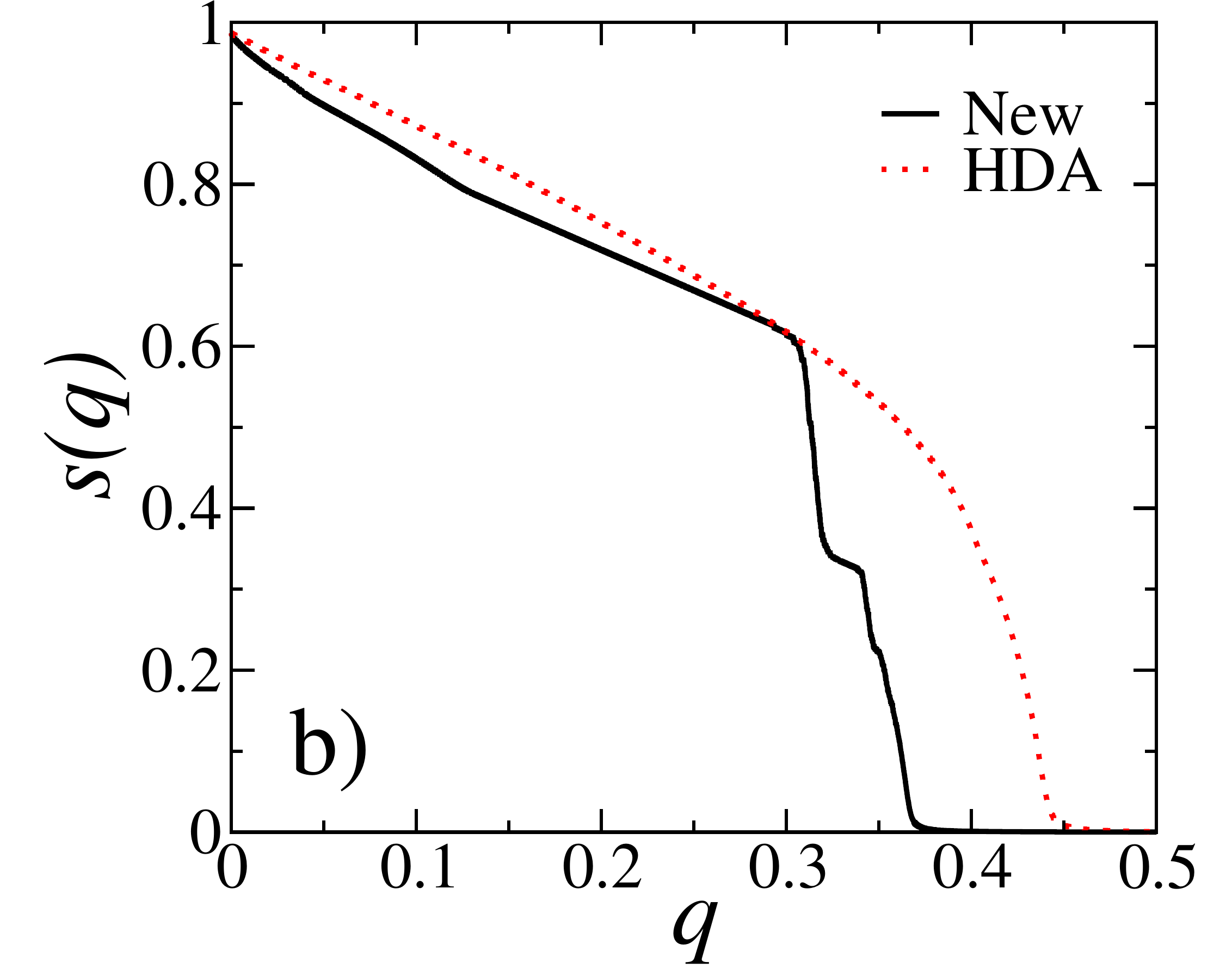}
\caption{The fraction $s(q)$ of sites belonging to the largest connected cluster versus the fraction $q = Q/N$ of immunized nodes using the new immunization strategy (full lines) and HDA attacks (dashed red lines) for (a) the AS Internet with $N = 18124$ and $M = 37357$ and (b) the HEP network with $N = 27240$ and $M = 341923$~. }
\label{fig:robustnessReal}
\end{figure}
\section{Results}
We compare the effectiveness of our inverse targeting immunization strategy to HDA and HBA strategies. We start by comparing the fractions $q_c$ of immunized nodes for which the network of non-immunized individuals breaks apart, which is the measure usually used in studies of epidemics. In addition, we use network's robustness to immunization as the measure of how effective the immunization procedure is. A good immunization strategy should make such a robustness as small as possible. The robustness sums up the sizes of the largest connected clusters $S(Q)$ of the networks of non immunized nodes remaining after immunization of Q nodes:
\begin{eqnarray*}
R = \frac{1}{(N + 1)} \sum_{Q = 0}^{N} S(Q)
\end{eqnarray*}
where $N$ is the size of the network. This measure captures the network response to immunization throughout the immunization process, and not only at the percolation point at which the network of non immunized nodes becomes disconnected \cite{christianPNAS, christianJstat}.                                                                                    
The process of immunization usually takes some time. The initially slow production of needed vaccine soon after the disease starts spreading, slows down the immunization process in these initial phases. Therefore it is important that the immunization process not only completely breaks up the network of contacts among susceptible, non-immunized individuals, but also that the size of the largest connected cluster, that is, the number of individuals being at highest risk of getting infected, is as small as possible during the whole immunization process. This property is expressed by a small robustness $R$.

We have applied our strategy together with HDA and HBA strategies to three different model networks, so called random regular graphs \cite{randomRegular} (in which nodes are connected randomly keeping the number of links per node constant), Erd\H{o}s-R\'enyi networks \cite{Erdos} and scale free networks created with the configuration model \cite{CM}. In Figure \ref{fig:robustness} we show how the fraction of nodes in the largest connected non-immunized cluster $s(q)$ changes with the fraction $q$ of immunized nodes during different immunization processes on these three model networks. $q_c$ is the value for which the largest connected cluster is of order $1/N$, and the robustness $R$ is the area under the curves in Figure \ref{fig:robustness}.

We can see that for all three types of model networks our immunization strategy has a smaller $q_c$ than HDA and has approximately the same value as for the HBA immunization. At the same time the robustness of networks in the case of our new strategy is lower than in other strategies making it the most efficient of the three strategies.

We have also applied our strategy to two real networks, the Internet at the level of autonomous system \cite{ASinternet}, as example for a network through which computer viruses can spread, and the collaboration network of high energy physicist (HEP) as an example of social networks \cite{HEP}. Here we could compare our strategy only to the HDA immunization strategy since HBA would take too much computer time. The results are presented in Figure \ref{fig:robustnessReal}, where we can see that in both cases the new immunization strategy is according to both criteria, $q_c$ and $R$, much more efficient than the HDA strategy.

Summing up the results presented in Figures \ref{fig:robustness} and \ref{fig:robustnessReal}, we have found that our immunization strategy outperforms the other two. Comparing to the HDA (HBA) strategy, in the case of the scale free network, $q_c$ is $17(0)\%$ and $R$ $11(4)\%$ smaller, for the Erd\H{o}s-R\'enyi network, $q_c$ is $24(0)\%$ and $R$ $11(7)\%$ smaller, and for the random regular network $q_c$ is $16(2)\%$ and $R$ $14(8)\%$ smaller. For real networks, for HEP network, $q_c$ is $18\%$ and $R$ $16\%$ smaller and for the AS Internet, $q_c$ is $37\%$ and $R$ $33\%$ smaller.
\begin{figure}[t]\centering
\includegraphics[height=6cm, angle = -90]{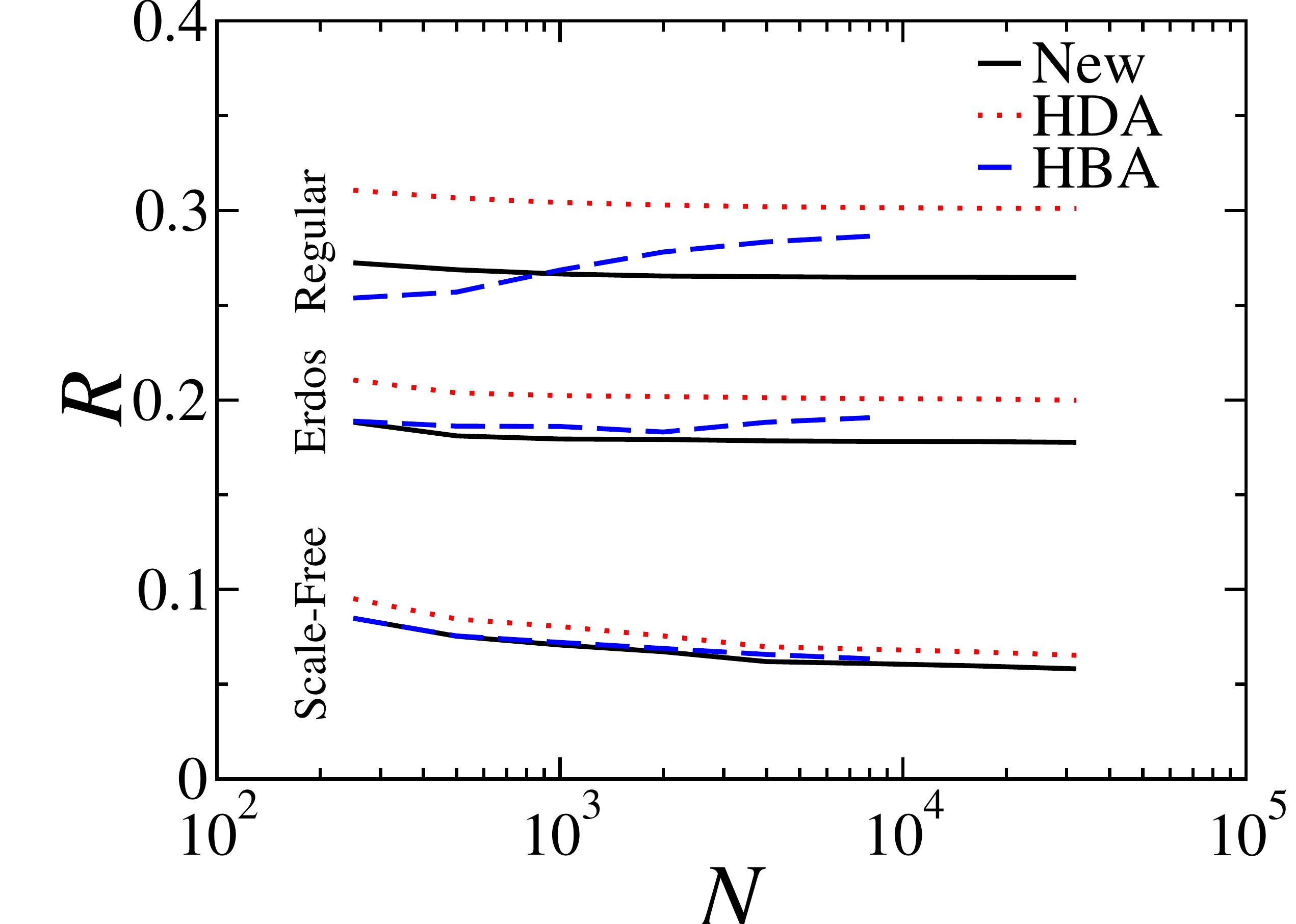}
\caption{The robustness versus the system size $N$ for random regular networks with $\langle k \rangle = 4$, Erd\H{o}s-R\'enyi networks with $\langle k \rangle = 4$, and scale-free networks with $\gamma = 2.5$ and $k_{\mathrm{min}} = 2$.}
\label{fig:DDA}
\end{figure}

For model networks we have studied the effect of the system size on the efficiency of our immunization strategy and have compared it with the other two strategies. In Figure \ref{fig:DDA} we can see that our immunization strategy is for all network sizes more effective than HDA (gives the smallest $R$ value). For Erd\H{o}s-R\'enyi and scale free networks it is always more efficient than the HBA strategy and only for small regular networks with less than 1000 nodes HBA is more efficient.

\begin{figure}[t]\centering
\includegraphics[height=5.5cm, angle= -90]{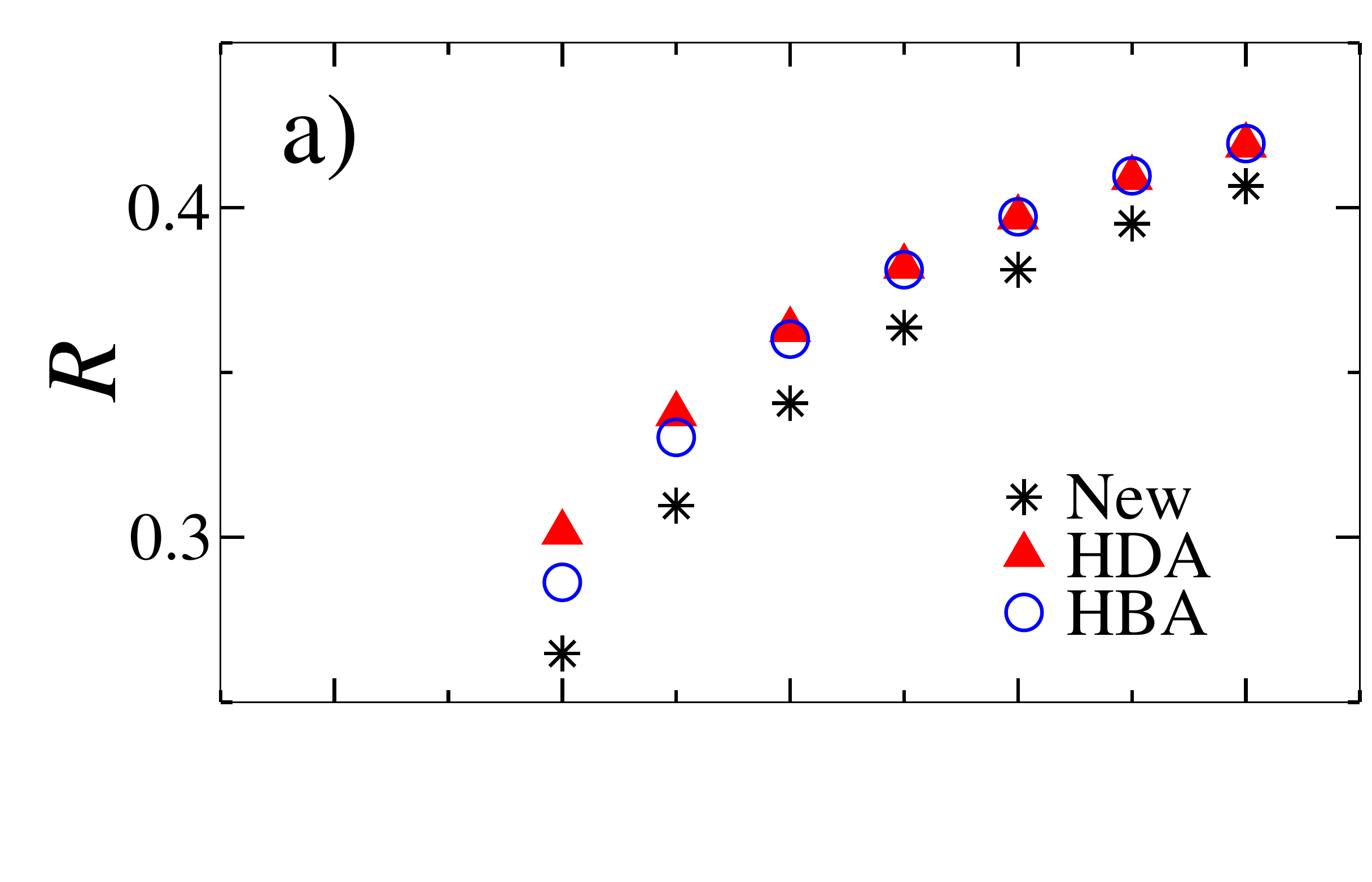}\\
\vspace*{-15px}
\includegraphics[height=5.5cm, angle= -90]{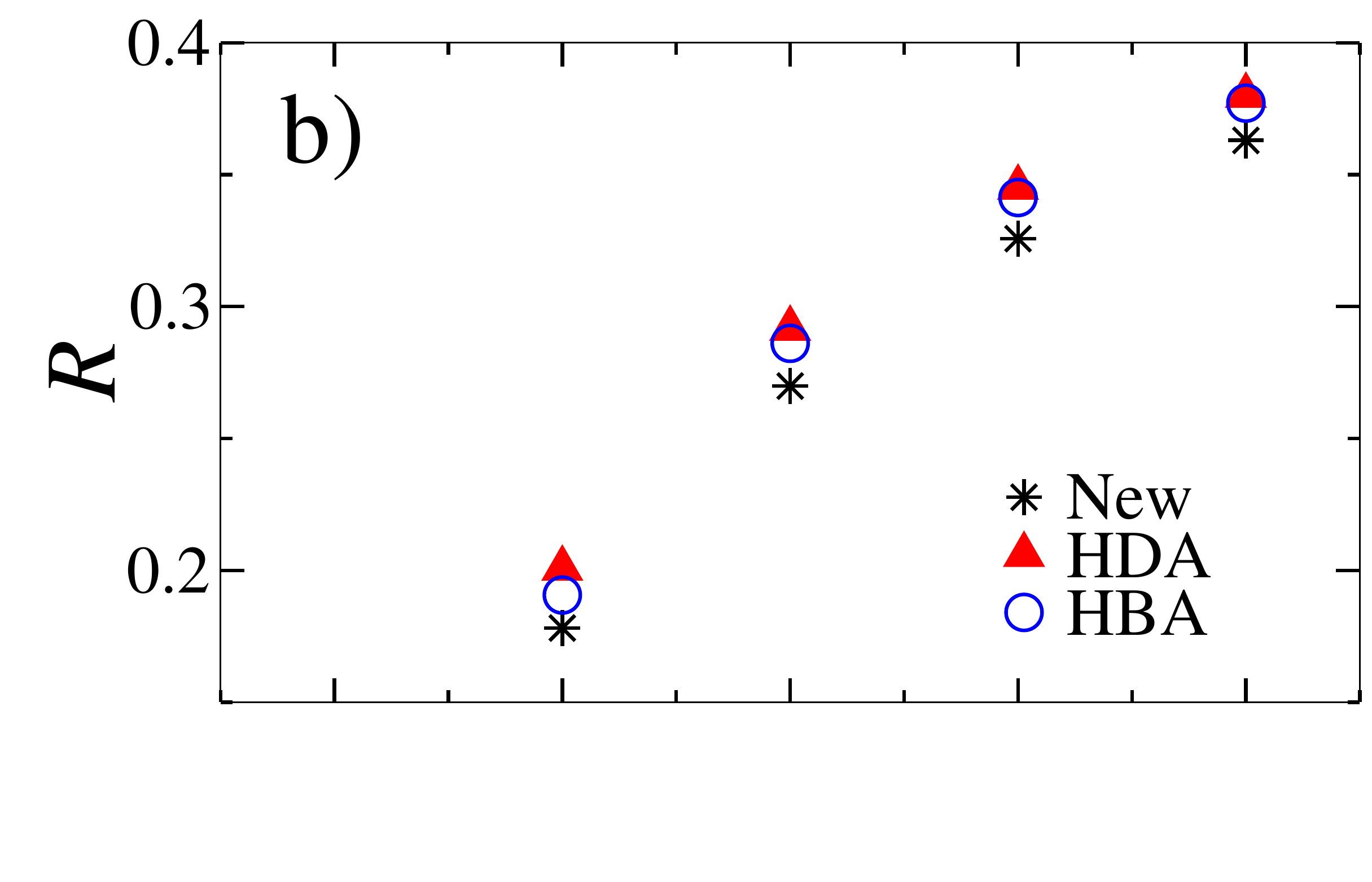}\\
\vspace*{-15px}
\includegraphics[height=5.5cm, angle= -90]{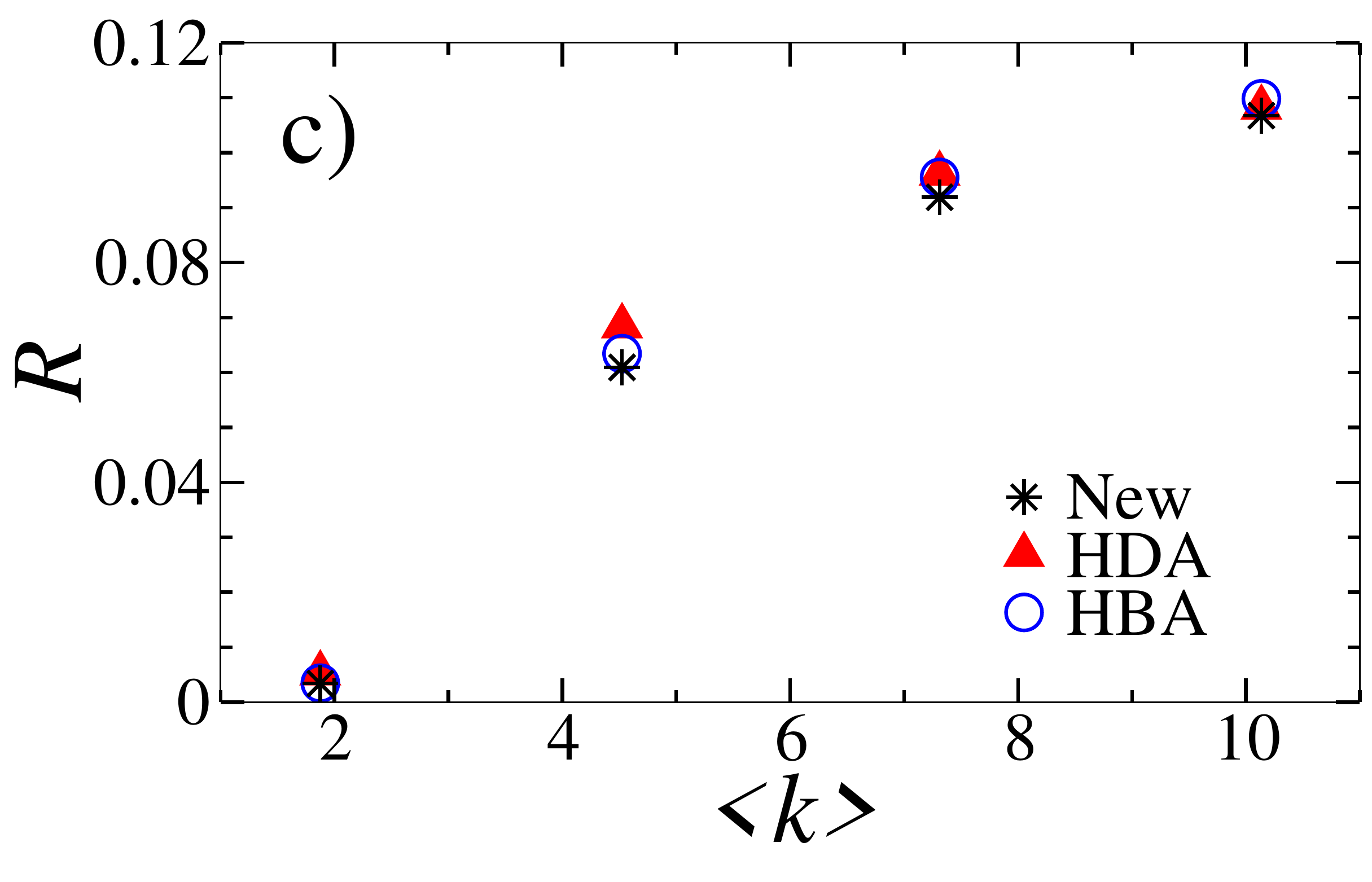}
\caption{The robustness $R$ for all three strategies versus average degree $\langle k \rangle$ for (a) random regular, (b) Erd\H{o}s- R\'enyi, and (c) scale-free networks with $N = 8000$.}
\label{fig:threshold}
\end{figure}

The way the effectiveness of immunization strategies changes with the density of links in the model networks is shown in Figure \ref{fig:threshold}. We can see that the new immunization strategy is the most efficient for all average degrees $\langle k\rangle$ studied. In all cases studied here (for different $\langle k\rangle$ and different $N$) the difference in the effectiveness of different strategies is smallest for scale free networks.
\section{Conclusions}
We have introduced a novel immunization strategy based on inverse targeting. To estimate the efficiency of our strategy we have not only used the percolation transition $q_c$ but also an additional measure taking into account the history of the immunization process. We find that the new strategy is much more efficient than the high degree based adaptive targeted strategy. It is also at least as efficient as the high betweenness based targeting strategy which was for a long time claimed to be the most effective immunization strategy \cite{holmeAttacks}. Due to its numerical inefficiency, the HBA strategy can not be used for larger systems and is therefore not applicable to real world problems. Our strategy is in contrast also numerically efficient and can therefore become the strategy of choice for real applications. The algorithm we introduced could find other applications besides immunization. It could for example be used to determine attack strategies for the efficient destruction of criminal networks.

\section{Acknowledgments}
We acknowledge financial support from the ETH Competence Center 'Coping with Crises in Complex Socio-Economic Systems' (CCSS) through ETH Research Grant CH1-01-08-2 and the Swiss National Science Foundation under contract 200021 126853, the Swiss National Science Foundation under contract 200021 126853 and FUNCAP.

\end{document}